\documentclass[twocolumn,showpacs,preprintnumbers,amsmath,amssymb,aps,prb,floatfix,groupedaddress]{revtex4}
\usepackage{graphicx} 
\usepackage{dcolumn} 
\usepackage{bm} 
\usepackage{epsfig}
\usepackage{color}
\usepackage{multirow}

\begin{document} 

\title{NaAlSi: a self-doped semimetallic superconductor \\with free electrons
and covalent holes}

\author{H. B. Rhee}
\affiliation{Department of Physics, University of California, Davis, CA 95616}

\author{S. Banerjee}
\affiliation{Department of Physics, University of California, Davis, CA 95616}

\author{E. R. Ylvisaker}
\affiliation{Department of Physics, University of California, Davis, CA 95616}

\author{W. E. Pickett}
\affiliation{Department of Physics, University of California, Davis, CA 95616}

\date{\today}

\begin{abstract}
The layered ternary $sp$ conductor NaAlSi, possessing the iron-pnictide ``111'' 
crystal structure, superconducts at 7 K. Using density functional methods, we
show that this compound is an intrinsic (self-doped) low-carrier-density 
semimetal with a number of unusual features.  Covalent Al-Si valence bands
provide the holes, and free-electron-like Al $3s$ bands, which propagate in
the channel between the neighboring Si layers, dip just below the Fermi level to
create the electron carriers. The Fermi level (and therefore the superconducting
carriers) lies in a narrow and sharp peak within a pseudogap in 
the density of states. The small peak
arises from valence bands which are nearly of pure Si, quasi-two-dimensional, flat,
and coupled to Al conduction bands. Isostructural NaAlGe, which is not
superconducting above 1.6 K, has almost exactly the
same band structure except for one missing piece of small Fermi surface.  Certain
deformation potentials induced by Si and Na displacements along the $c$-axis are
calculated and discussed. It seems likely that the mechanism of pairing is
related to that of several other lightly doped two-dimensional nonmagnetic
semiconductors (TiNCl, ZrNCl, HfNCl), which is not well understood but
apparently not of phonon origin.
\end{abstract}

\maketitle

\section{Introduction}
The discovery of new superconductors in unexpected materials brings
the potential to understand something deeper, or perhaps something
different, about the underlying properties that
favor superconducting pairing.  The discovery of high-temperature
superconductivity (to 56 K so far) in Fe-pnictides\cite{hosono} is a recent
spectacular example, and is also an example of close relationships
between magnetism and superconductivity, though the connections
are still far from clear.

New superconductors where little or no magnetic effects are present are
also arising, and these clearly involve different physics from
the cuprate or Fe-pnictide high-temperature superconductors. Electron-doped
MNCl, where M = Zr or Hf, becomes superconducting immediately
upon undergoing the insulator-to-metal
transition,\cite{zrncl1,zrncl2,hfncl1,hfncl2} which, in the case of
M = Hf, is higher than 25 K.  The similarly
layered, electron-doped, ionic insulator TiNCl superconducts up to
16 K.  Magnetic behavior in these materials is at most subtle,
amounting to an enhancement in Pauli susceptibility near the
metal-to-insulator transition.\cite{zrncl3}

In this paper we address the ternary silicide NaAlSi (space group \emph{P4/nmm},
$Z=2$), another ionic and layered material that shows unexpected
superconductivity, and does so in its native (without
doping or pressure) stoichiometric state, at 7
K.\cite{kuroiwa} NaAlSi introduces new interest from several viewpoints. First,
it is an $sp$ electron superconductor, with a high $T_c$ for such materials
at ambient pressure. Pb is an $sp$ superconductor with comparable $T_c$ (7.2 K)
but with simple metallic bonding and heavy atoms, making it very different. 
Doped Si\cite{silicon} and doped diamond\cite{diamond} superconduct 
in the same range, and are of course very different classes of materials.  A
more relevant example is the pseudo-ternary compound Ba$_{1-x}$K$_x$BiO$_3$ (BKBO), which
undergoes an
insulator-to-metal transition for $x \approx 0.4$, beyond which its $T_c$
surpasses 30 K.[\onlinecite{bkbo,bkbo2}]

Second, the Al-Si layered substructure is like that of the FeAs layer in the
Fe-pnictide superconductors, raising the possibility of some connections
between their electronic structures. In fact, NaAlSi has the same structure as 
the
Fe-pnictide ``111" compounds, with Al being tetrahedrally coordinated by Si
(analogous to Fe being tetrahedrally coordinated by As).  In spite of their
structural similarities, these compounds have major differences; for example,
the Fe pnictides are $3d$ electron systems with magnetism, while NaAlSi
is an $sp$ electron system without magnetism.

Third, NaAlSi is the isovalent sister (one
row down in the periodic table for each atom) of LiBC. LiBC itself is (in a
sense) isovalent and also isostructural to MgB$_2$; however due to the B-C
alternation around the hexagon in the honeycomb-structure layer, LiBC is
insulating rather than conducting. When hole-doped while retaining the same
structure, Li$_{1-x}$BC has stronger electron-phonon coupling than does
MgB$_2$.\cite{rosner} While NaAlSi has a substantially different structure than
LiBC, its isovalence and its combination of covalence with some ionic character is
shared with LiBC.

Yet another closely related compound is CaAlSi, whose two different stacking
polymorphs and parent structure superconduct in the 5--8 K
range.\cite{sagayama,kuroiwa2} Linear-response and frozen-mode
calculations indicate
electron-phonon coupling is the likely mechanism; in particular, an ultra-soft
phonon mode appears and is suggested to play a role in the 
superconductivity.\cite{shein, huang,
mazin, giantomassi, heid, kuroiwa3} It is curious that
in this compound, where divalent Ca (comparing it with NaAlSi) contributes one
additional electron into the
Si-Al $sp$ bands, the preferred structure is that of AlB$_2$ (\textit {i.e.},
MgB$_2$) with $sp^2$ planar bonding\cite{shein, huang, giantomassi} rather than
the more $sp^3$-like bonding in NaAlSi. Electronic structure calculations show
that CaAlSi has one electron in the conduction band above a bonding-antibonding
band separation at the NaAlSi band-filling level, a situation which would 
not appear to be particularly
favorable for $sp^2$ bonding.

In this paper we analyze first-principles electronic structure calculations that
reveal that NaAlSi is a naturally self-doped semimetal, with doping
occurring---thus charge transfer occurring---between covalent bands
within the Al-Si substructure, and two-dimensional (2D) free-electron-like bands
within the Al layer. The resulting small Fermi surfaces (FSs) are
unusual, complicated by the small but seemingly important interlayer coupling
along the crystalline $c$-axis.

\begin{figure}[b]
\begin{center} 
\includegraphics[draft=false,width=\columnwidth]{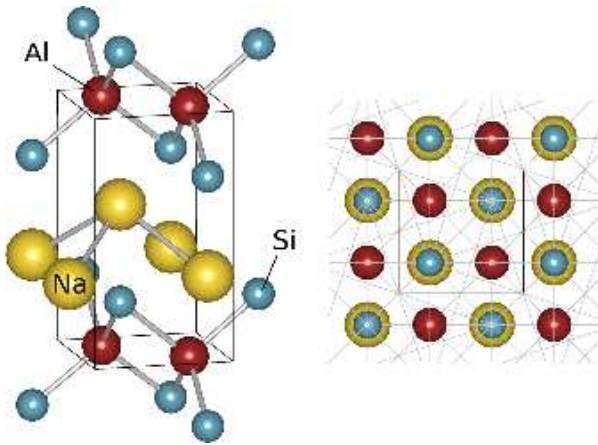}
\end{center} 
\caption{(Color online.) Crystal structure\cite{vesta} of NaAlSi. Four Si atoms
tetrahedrally surround an Al atom, and these Al-Si networks sandwich the Na
atoms. The unit cell is outlined in black.}
\label{structure}
\end{figure}

\section{Computational Methods}
First-principles, local density approximation (LDA) calculations were carried
out using the full-potential local-orbital (FPLO) scheme.\cite{fplo} A $k$-point
mesh of $20 \times 20 \times 12$ was used, and the Perdew-Wang 92
approximation\cite{pw92} was applied for the exchange-correlation potential. The
experimental lattice constants obtained by Kuroiwa \textit{et
al.}\cite{kuroiwa} ($a = 4.119$~\AA\ and $c = 7.632$~\AA) and internal
coordinates published by Westerhaus and Schuster\cite{westerhaus} ($z_{Na}=0.622$,
$z_{Si}=0.223$) were used in our calculations.

\section{Electronic Structure}
\subsection{Discussion of the band structure}

\begin{figure}[b]
\begin{center} 
\includegraphics[draft=false,width=\columnwidth]{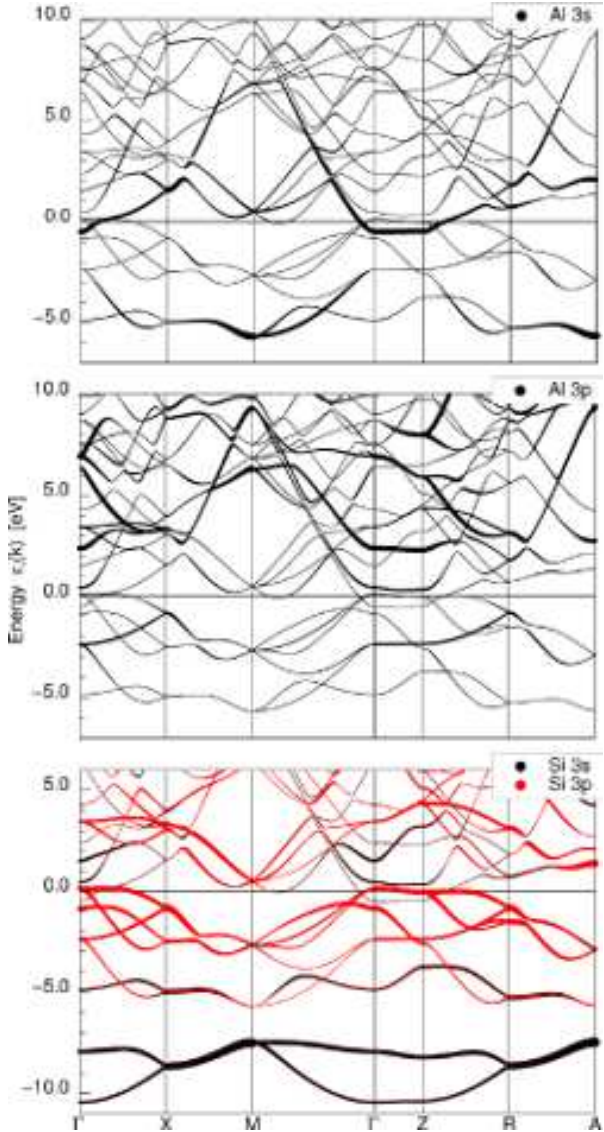}
\end{center}
\caption{(Color online.) Band structure, with projected fatbands, of NaAlSi.
Top panel: the Al 3$s$ 
character of bands is indicated by broadening.  A doubly degenerate pair
of broad bands is evident along the $\Gamma$-M direction.  Middle panel:
Al $3p$ character is weak below 2--3 eV.  Bottom panel: Si $3s$ (black) 
bands below $-8$ eV, and Si $3p$ in the valence bands and lower conduction
bands.}
\label{bs}
\end{figure}

The calculated band structure of NaAlSi is shown in Fig.\ \ref{bs}. As expected,
the Na ion gives up its electron to the Si-Al-Si trilayer (see
Fig.\ \ref{structure}), which may have some ionic character, though it is not
easy to quantify (the valence bands have much more Si character than
Al character, seemingly more than suggested by their number of valence electrons).
There are several readily identifiable classes of bands. Two primarily Si $3s$
bands, with a small amount of Al $3s$ character, are centered 9 eV below the
Fermi level $\varepsilon_F$ and have a width of 2.5 eV. Above them there is a
six-band complex of Al-Si $s$-$p$ bands (heavily Si) that are very nearly
filled, the band maximum only slightly overlapping $\varepsilon_F$.

Above $\varepsilon_F$ lie non-bonding and antibonding bands, and the Na $s$
bands.  Among these there are a pair of distinctive bands, which can be identified
most easily by their Al $s$ character in the top panel of Fig.\ \ref{bs}. These
bands are nearly free-electron-like with large dispersions, and cross many
other bands with little mixing. Along $\Gamma$-M
they are degenerate and easily identifiable in Fig.\ \ref{bs}, as they disperse
up through the Fermi level to nearly 10 eV at the M point.  Along $\Gamma$-X,
and similarly at the top of the zone Z-R, they are distinct: one again disperses
upward rapidly, cutting through many other bands, also to nearly 10 eV at X; the other
disperses much more weakly to X, with a bandwidth of about 2 eV.  Their Al $s$
character and nearly vanishing Si character identify these as free-electron
states, in which electrons move down channels of Al atoms separately in $x$
and $y$ directions. (Note that their lack of $k_z$ dispersion identifies them as
planar bands.) There is some coupling to the states in a parallel channel of Al
atoms, giving rise to the 2 eV transverse dispersion.  These bands lie 0.5 eV
below $\varepsilon_F$ at $\Gamma$ and contain electrons.  Without interference
with other bands near the Fermi level and supposing them to be isotropic in the
plane (but see below), such FSs might include 3--4\% of the area of the zone,
which would equate to an intrinsic electron doping for two bands, both spins of
around 0.12--0.16 carriers per unit cell, and the concentration of hole carriers would be
equal.  The anisotropy, discussed below, makes the actual carrier concentration
much lower.

The small overlap of valence and conduction bands results in semimetallic
character and small Fermi surfaces. The valence bands are quite anisotropic.  
Looking at the valence bands along
$\Gamma$-X, one might try to characterize them as one ``heavy-hole'' and one
``light-hole'' band, degenerate at $\Gamma$, with the band maximum lying 0.13~eV 
above $\varepsilon_F$.  However, the heavy hole band is actually almost perfectly flat for
the first third of the $\Gamma$-X line, before dispersing downward
across $\varepsilon_F$ and farther below. Due to this flatness, the band cannot
be characterized by an effective mass.
The conduction bands contribute the pair of light electron bands described above.
In addition, one conduction band dips slightly below the Fermi level along
$\Gamma$-M near M.

The fatbands representation in Fig.\ \ref{bs} that reveals the dominant band
character shows that the Si $3p_x$, $3p_y$, and Al $3s$ orbitals dominate the
valence states near $\varepsilon_F$. As anticipated from consideration of the
layered structure as mentioned earlier, the electronic structure is quasi-2D,
with generally small dispersion along $k_z$ near $\varepsilon_F$. 
However, the small $k_z$ dispersion of
one band is important in determining the
geometry of the FSs, as discussed in more detail below.

\begin{figure}[b]
\begin{center} 
\includegraphics[draft=false,width=\columnwidth]{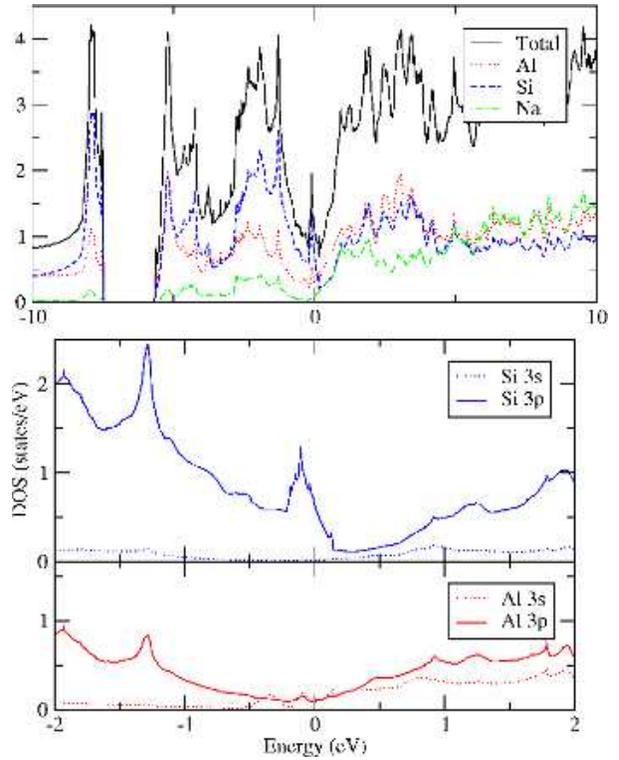}
\end{center} 
\caption{(Color online.) Total and partial (atom- and orbital-projected)
DOSs of NaAlSi.
Top panel: Total and atom-projected DOS in 
a 20 eV-wide region, showing the pseudogap centered at the Fermi level (the
zero of energy) punctuated by the curiously narrow and sharp peak at the Fermi
level. Middle panel: expanded view of the peak, and the variation of the DOS 
near the Fermi level, separated into Si $s$ and $p$ contributions. Lower panel:
the Al $s$ and $p$ character; the $s$ character ``turns on'' just below the
Fermi level.}
\label{dos}
\end{figure}  

\subsection{Density of states}
Fig.\ \ref{dos} shows the total, partial, and projected densities of states
(DOS) of NaAlSi. The Na contribution near the Fermi level is negligible and thus
not shown. Except for a strong dip (``pseudogap'') near the Fermi level 
and a less severe dip in the
$-3$ to $-4$ eV range, the DOS hovers around 3 states/eV throughout both valence
and conduction bands. Within the pseudogap encompassing the Fermi
energy, there is an anomalous sharp and narrow peak with $\varepsilon_F$
lying on its upper slope, as noted
previously by
Kuroiwa \textit{et al.}\cite{kuroiwa}  The value of $N(\varepsilon_F)$ is 1.1
states/eV.  We discuss below the FSs of both hole (Si) and electron
(Al) character.

It seems clear that the transport properties and low-energy properties (which
have not yet been reported), and in
particular the superconductivity of NaAlSi, are intimately associated with this
sharp and narrow peak in the DOS, which includes the Fermi level. The projected
DOS shows the flat bands that give rise to this peak are very strongly Si-derived.
There is Al $3s$ character that turns on just below $\varepsilon_F$, but it is 
relatively small compared to the Si
character at $\varepsilon_F$, and its magnitude remains low and nearly constant through the peak.
There is Al $3p$ character of the same magnitude in the vicinity of the Fermi level.
The top edge of the peak coincides with the flat band along $\Gamma$-X at 0.13 
eV. The width of the peak, about 0.35 eV, must be due to dispersion and
anticrossings that are mostly not visible along symmetry directions and arise
from mixing away from symmetry lines of the valence and conduction bands.

Nonetheless, the slope of the DOS at $\varepsilon_F$ is rather steep, and this
may give rise to high thermopower for the material. 
The standard low-temperature
limit of thermopower (the Seebeck coefficient tensor) {\bf S}$(T)$ in
semiclassical Bloch-Boltzmann theory is
\begin{eqnarray}
{\bf S}(T) \rightarrow -\frac{\pi^2 k_B}{3e} \left . \frac{d \ln {\boldsymbol \sigma}(E)}
{dE}\right \vert _{\varepsilon _F} k_B T.
\label{seebeck}
\end{eqnarray}
The conductivity tensor ${\boldsymbol \sigma}(E)$ can be written in terms of the
average velocity ($\vec v(E)$) product, DOS, and scattering time
$\tau(E)$ over the constant energy ($E$) surface:
\begin{eqnarray}
\boldsymbol \sigma(E) = 4\pi e^2 \langle \vec v(E) \vec v(E)\rangle N(E)\tau(E).
\end{eqnarray}
The thermopower thus picks up contributions from the energy variation of three
quantities: the dyadic product
$\langle\vec v \vec v\rangle$, $N(E)$, and $\tau(E)$. Often the energy dependence of $\tau$
is neglected, out of lack of detailed knowledge, though it also can
be argued to follow roughly $1/\tau(E) \propto N(E)$ for elastic scattering. 
The energy dependence of
$v^2(E)$ also counteracts the energy dependence of $N(E)$. Nevertheless it is
observed that materials with large slope in $N(E)$ frequently have large thermopower.
For NaAlSi we calculate $d \ln N(E)/
dE\vert _{\varepsilon _F}=-4.0$ eV$^{-1}$.  This value can be compared with other
materials that have fine structure near the Fermi level: TiBe$_2$, where 
$d \ln N(E)/
dE\vert _{\varepsilon _F}$ = 10--12 eV$^{-1}$ and $N(\varepsilon_F)$ also is much 
larger;\cite{kyker} and
MgCNi$_3$ with its very impressive peak very near $\varepsilon_F$, for which
$d \ln N(E)/
dE\vert _{\varepsilon _F}$ $\sim$ -15-20/eV$^{-1}$.\cite{rosner2}

The energy derivative of diagonal elements of $\boldsymbol\sigma$ that occurs in Eq.\ \ref{seebeck}
can also be expressed as 
\begin{eqnarray}
\frac{1}{\sigma}\frac{d\sigma}{dE}&=&\frac{d \ln \tau(E)}{dE} 
\nonumber \\
   & &+ \frac{1}{2\pi^2}  \frac{{\cal M}^{-1}(E)}{v^2(E)},
\end{eqnarray}
where ${\cal M}^{-1}(E)$ is (a diagonal element of) the inverse mass tensor (second derivative of
$\varepsilon_k$) averaged over the constant energy surface.  This form makes it
clear that the expressions should, for any quantitative estimate, be generalized
to two-band form, since the valence and conduction bands have differing signs of
their effective masses, and the scattering time---and its energy variation---is
likely to be very different for Si-derived covalent valence bands and Al-derived
free-electron conduction bands. Measurement of the thermopower, and a
quantitative theoretical treatment, would be very useful in extending the
understanding of the transport properties of NaAlSi.

\begin{figure}[b]
\begin{center} 
\includegraphics[draft=false,width=\columnwidth]{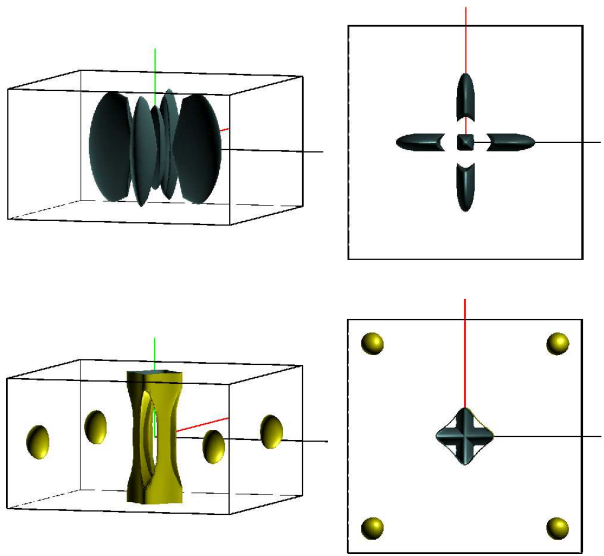}
\end{center} 
\caption{(Color online.) Views from the $xy$-plane (left) and top (right) of the
FSs of NaAlSi, centered at $\Gamma$. The blue (dark) surfaces enclose holes and
the yellow (light) surfaces enclose electrons.}
\label{fs}
\end{figure}

\subsection{Unusual Fermi surfaces}
Fig.\ \ref{fs} depicts the calculated FSs. In spite of the generally 2D band
structure, the small $k_z$ dispersion of bands at $\varepsilon_F$ make some of
the FSs surprisingly three-dimensional. Electron pockets and hole
pockets coexist in the Brillouin zone, with electron and hole concentrations
necessarily being equal. 

{\it Hole surfaces.} Four hole ``fan-blade'' surfaces lie oriented in the $xz$-
and $yz$-planes.  At the center, extending from $\Gamma$ half way to Z, lies a
long and narrow surface with square cross section. The top view allows the
origin of these surfaces to be understood.  The cross sections in the $xy$-plane
are of two ellipses that are very anisotropic (in the $xy$-plane) and at right
angels to each other. Each corresponds to a dispersion that is weak in one
direction (the long major axis) and strong in the other (minor axis). These
bands would intersect, but in fact are intersected by the electron band that
cuts a squarish hole (rotated by 45$^{\circ}$), within which the elongated hole
surface inside re-emerges. 

{\it Electron surfaces.} In the bottom panels of Fig.\ \ref{fs}, the squarish
electron surface (with $k_z$ variation and resulting holes, shown in the lower
two panels of Fig. \ref{fs}) that cuts the
aforementioned hole surface is pictured, and substantiates the discussion
provided just above.  In addition, there are simple electron ellipsoids centered
along the $\Gamma$-M lines.  It is curious that in a band structure that is for
the most part strongly 2D, all the FSs have a rather definite three-dimensional
character. Although the bands show little dispersion along $\Gamma$-Z, the
bands just above the the Fermi level are quite different depending on whether
$k_z = 0$ or $k_z = \pi$.  In particular, the lowest band along R-A is rather
flat, but the lowest conduction band along X-M has a dispersion of nearly 2 eV.
Similar comparisons can be made for the bands along $\Gamma$-X and Z-R. The
$k_z$ dispersion is not nearly as strong near $k_x = k_y = 0$, which is clear
from both the band structure and the FS.

{\it Short discussion.} It was noted in the Introduction that the 
NaAlSi structure is the same as the
Fe-pnictide ``111'' structure.  Moreover, in both compounds, the relevant bands
involve only the (Si-Al-Si or As-Fe-As) trilayer.  The top view of the 
fan-blade surfaces have characteristics in common with those of some of
the Fe pnictides,\cite{singh, mazin2} all of which have this same trilayer.
The similarity is that the top view of the fan blades (if one ignores the
diamond-shaped cutout at the intersection, centered at $\Gamma$) appears to show
intersecting FSs, neither of which has the square symmetry of the lattice.

Such occurrence of intersecting FSs, each with lower symmetry than
the lattice, has been analyzed for LaFeAsO (a ``1111'' compound) by 
Yaresko {\it et al.}\cite{yaresko}
A symmetry of the Fe$_2$As$_2$ (also Al$_2$Si$_2$)
substructure is a non-primitive translation connecting Fe atoms (respectively,
Al atoms) followed by $z$-reflection.  This operation leads to symmetries
that allow $k_z = 0$ bands to be unfolded into a larger Brillouin zone (that is,
a ``smaller unit cell'' having only one Fe atom) which {\it un}folds the
band structure and the intersecting FSs.  The NaAlSi FSs
appear to have this similar crossing (albeit interrupted by the free-electron bands), and
the highly anisotropic dispersion is due to distinct (but symmetry-related) hopping along 
each of the
crystal axes.
In this respect  NaAlSi may clarify the electronic structure of the pnictides:
by analogy, there are separate bands that disperse more strongly along the $(1,1)$ direction or
the $(1,-1)$ direction, and give rise to the intersecting, symmetry-related 
surfaces.  In NaAlSi the bands are much more anisotropic in the plane 
(approaching one-dimensional), making such character much clearer.
A difference that complicates the analogy is that in the pnictides the bands near
$\varepsilon_F$ are derived from the Fe atoms, which comprise the center layer
of the trilayer, whereas in NaAlSi the bands under discussion
are derived from the Si
atoms, which comprise the two outer layers.

\subsection{Wannier functions}

\begin{figure}[b]
\begin{center}
\includegraphics[width=\columnwidth]{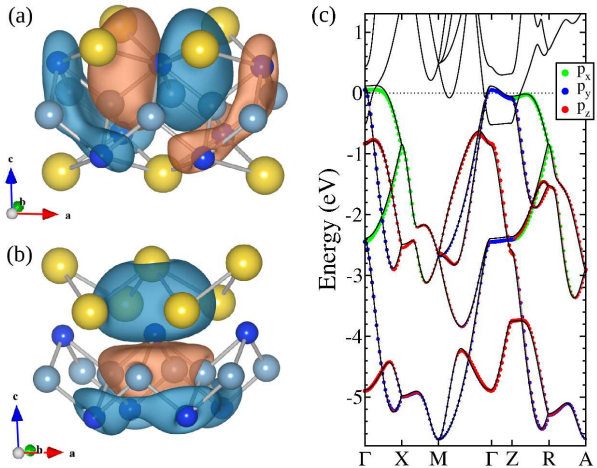}
\caption{(Color online.) Isosurface of the WFs for (a) Si $3p_x$
and (b) Si $3p_z$. Na atoms are large and yellow (light) colored, Si atoms are
small and blue (dark) colored.  The two colors of the isosurface represent
different signs. (c) The tight-binding fatbands band structure described in the
text for the WFs, compared to the DFT band structure (black lines).}
\label{fig:Wannier}
\end{center}
\end{figure}

Pictured in Fig.\ \ref{fig:Wannier} are symmetry-projected Wannier functions
(WFs) projected onto Si $3p$ orbitals. The extension of the WFs shows
considerable involvement from nearby Al and Si atoms, and in addition
have some density extending into the Na layers.
The $p_x$ WF consists of an atomic $p_x$ function, with its density
shifted downward by the bonding contribution of Al $sp^3$ hybrid orbitals.
Beyond the $p_x$ lobes the nearest Si atoms form a bonding lobe that connects to
the ``small'' side of the Al $sp^3$ function. The large $p_x$ lobes and the extra
contribution from nearby Si atoms are responsible for the largest hopping amplitudes shown
in Table \ref{tbl:TightBinding}, although there is some phase cancellation
between the $p_x$ lobe and the lobe lying beyond the nodal surface.

The $p_z$ WF has one lobe extended well into the Na layer; this is
responsible for the largest hoppings along $\mathbf{b}^*$ in Table
\ref{tbl:TightBinding}, and they create the large dispersion in the $p_z$ bands
seen in Fig.\ \ref{fig:Wannier}(c).  Again, the Al atoms contribute with an $sp$
hybrid orbital, although it appears to be more $sp^2$-like than $sp^3$-like.
There is
also a ``ring'' structure below the Al layer, where an $sp$ hybrid orbital from
the Si atoms forms a bonding combination, but it is antibonding with the
$p_z$ function on the central Si. The largest contribution to near-neighbor
hopping in the Al-Si plane between $p_z$ and $p_x$ or $p_z$ is most likely due
to this ring structure, as the $p$ lobes are confined to the inside of a square
of near-neighbor Al atoms, which are only edge sharing with the nearest Si atoms
along $\mathbf{b}$ vectors. This is the likely reason that all the hoppings
along $\mathbf{b}$ are approximately of the same magnitude. The dispersion which
creates the FSs along $\Gamma$-Z (seen in Fig.\ \ref{fs}) is composed only of
the $p_x$ and $p_y$ WFs. This is not caused by the large hoppings, but by smaller
hoppings along $\mathbf{b}^*$ between $p_x$ and $p_y$ WFs.  Without these small
hoppings, the band just above $\varepsilon_F$ is dispersionless along $\Gamma$-Z.

\begin{table}[t]
\begin{center}
\begin{tabular}{c|cccc}
			&&  $p_x$    & $p_y$    & $p_z$        \\
\hline
\multirow{3}{*}{$\mathbf{a}$} 
	& $p_x$    &   761   &     	&  60  \\
   & $p_y$    &         & -62    &  60  \\
	& $p_z$    &    60   & 	60   	&  -40 \\
\hline
\multirow{1}{*}{$\mathbf{2a}$} 
	& $p_x$    &   128   &     	& 27    \\
\hline
\multirow{3}{*}{$\mathbf{b}$} 
	& $p_x$    &   361   & 300 	& 360  \\
	& $p_y$    &   300   & 361 	& 360  \\
	& $p_z$    &   360   & 360    & 360  \\
\hline
\multirow{3}{*}{$\mathbf{b}^*$}
   & $p_x$    &   12    &   5    &   50  \\
   & $p_y$    &    5    &  12    &   50  \\
	& $p_z$    &   50    &  50    &  185 
\end{tabular}
\caption{Selected hopping integrals in meV for the Si $3p$ WFs
along the vectors $\mathbf{a} = (a,0,0)$ (hopping within a Si layer),
$\mathbf{b} = (a/2,a/2,d)$ (hopping across an Al layer), and $\mathbf{b}^* =
(a/2,a/2,c-d)$ (hopping across a Na layer). $d$ is the distance in the $z$
direction between Si atoms above and below Al planes.}
\label{tbl:TightBinding}
\end{center}
\end{table}

\section{Response to changes}
\subsection{Electron-ion coupling}
A deformation potential $\cal{D}$ is the shift in an energy band with respect to
sublattice atomic displacement. One can freeze in phonon modes to calculate
deformation potentials, which at the FS are directly connected to
electron-phonon
matrix elements.\cite{khan}

\begin{figure}[t]
\begin{center}
\includegraphics[width=\columnwidth]{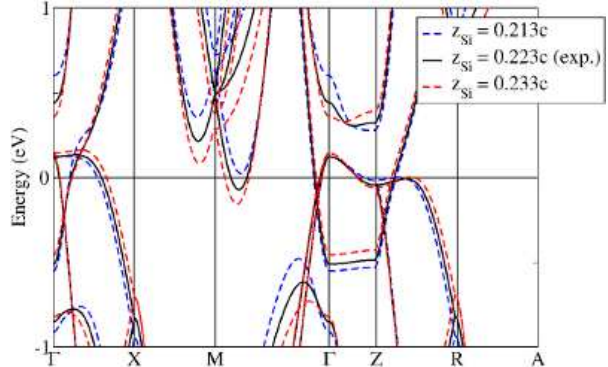}
\caption{(Color online.) Comparison of band structures near $\varepsilon_F$ for different
$z_{Si}$
values.}
\label{z-si}
\end{center}
\end{figure}

Moving the Si atoms in the $z$ direction by $\pm$1\% of the experimental parameter,
such that the tetrahedra surrounding the Al atoms stretch or flatten (while remaining
centered on Al), gives an average deformation potential of $\sim$0.8 eV/\AA\
over five band positions near the Fermi level.
The largest shift is for
the ellipsoidal electron pockets, with ${\cal D} \sim 1.2$ eV/\AA. These ellipsoids disappear
when the Si atoms are displaced toward the Al plane (see Fig.\ \ref{z-si}).

Flattening the Na bilayer, so as to remove the buckling of the Na atoms,
requires a (very large) 12\% change in
the $z$ component of the Na atoms. We chose such a large displacement because we do not
expect a substantial deformation potential for Na movement.
Even this large displacement does not alter very much
the valence bands, as expected, and the hole FSs remain virtually
unchanged. The conduction bands at the Fermi level however shift appreciably,
resulting in a modulation of the electron ellipsoids along (1,1) near M.
In addition, the accidental
four-band near-degeneracy that is 0.5 eV above $\varepsilon_F$ at
M splits the two separate doubly-degenerate states, opening up a gap of
$\sim$0.7 eV, which is equivalent to a deformation potential of $\sim$0.8
eV/\AA\ (but the bands are not at $\varepsilon_F$).

\subsection{Magnetic susceptibility}
The magnetic spin susceptibility $\chi$ is given by
\begin{equation*}
\chi = \frac{\partial M}{\partial H} = ({\partial^2 E /
\partial M^2})^{-1}.
\end{equation*}
Fixed-spin-moment calculations were conducted to produce an energy-vs.-moment
$E(M)$ curve, resulting in a susceptibility of $\chi = 2.93~\mu_B^2~$eV$^{-1}$,
or $3.76 \times 10^{-6}$ emu/mol.
This is the
exchange-enhanced susceptibility in the Stoner theory, and can be written as
\begin{equation*}
\chi = S \chi_0 \equiv \frac{\chi_0}{1 - I N(\varepsilon_F)},
\end{equation*}
where $S \equiv \chi/\chi_0$ is the Stoner enhancement factor, $I$ is the
Stoner parameter, and the bare Pauli susceptibility $\chi_0$ is equal to
$\mu_B^2 N(\varepsilon_F)$. Janak has shown how to calculate $I$ within
density functional theory,\cite{Janak} which must be equivalent (within a
minor approximation he used) to our approach  of using fixed-spin-moment
calculations.

With our calculated values we obtain $S = 2.75$, which translates to $I = 0.60$ eV.  Some
interpretation of  this value of Stoner $I$ should be noted.  First,
$\varepsilon_F$ falls where most of the states are
Si-derived, so for simplicity we neglect Al (and Na, which is ionized).  Second, we note that
there are two Si atoms in the primitive cell.  Thus to get an ``atomic value'' of
$I_{Si}$, we should use a value of $N(\varepsilon_F)/2$ per Si atom. 
The result then is
very roughly $I_{Si} \sim 1$ eV. This value can be compared to the atomic value for Al
(next to Si in the periodic table) in the elemental metal, which is $I_{Al}$ =
0.6 eV.\cite{Janak} It seems, therefore, that Si in NaAlSi is considerably more ``magnetically
inclined'' than is Al in aluminum.  However, the Stoner enhancement overall is 
not large, indicating relatively modest magnetic enhancement and rather conventional
magnitude of magnetic interactions.

\begin{figure}[b]
\begin{center} 
\includegraphics[draft=false,width=\columnwidth]{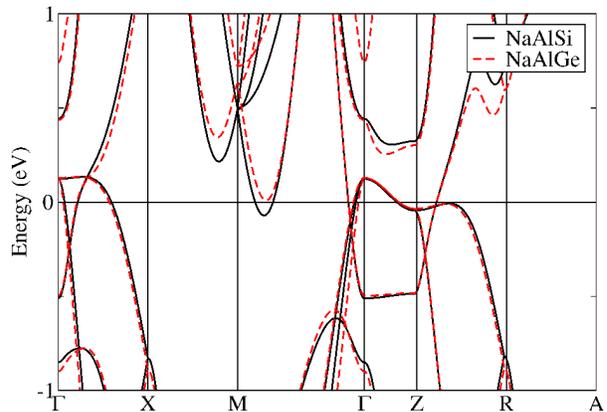}
\end{center} 
\caption{(Color online.) Blowup of the band structures of NaAlSi and NaAlGe near
$\varepsilon_F$.}
\label{naalge}
\end{figure}

\subsection{Comparison to NaAlGe}
Isostructural and isovalent NaAlGe is not superconducting (above 1.6 K, at
least), so it should be instructive to compare its electronic structure to that
of NaAlSi. Using its experimental lattice parameters, \cite{westerhaus} we have
calculated the band structure of NaAlGe, and compared it with that of NaAlSi on
a rather fine scale in Fig.\ \ref{naalge}. The band structures are very similar,
the one difference being that the band along $\Gamma$-M near M does not cross
$\varepsilon_F$ in NaAlGe. The free-electron band is also identical. 

Supposing the tiny bit of FS along $\Gamma$-M cannot account for the difference
in superconducting behaviors, the factors relevant for electron-phonon coupling
will be the difference in mass (Ge is more than twice as heavy as Si) and the
difference in electronic character, which can affect force constants and
electron-phonon matrix elements. A real possibility is that the pairing
mechanism is electronic rather than phononic. In three dimensions purely
electronic pairing mechanisms have been attracting serious study (by Sham
and collaborators\cite{sham1,sham2} for example), but 2D semimetals introduce new
features that deserve detailed study.

Another possibility is that these pockets are important, and that 
superconductivity arises from an enhancement of electron-phonon coupling
in these tiny electron pockets from low frequency electronic response,
either interband transitions or plasma oscillations, or both.  A model in 
which the electronic response of a 2D electronic superlattice plays a central
role in the 
mechanism has been previously studied\cite{bill1,bill2} using a model of
parallel conducting sheets separated by a dielectric spacer.  This model may
be useful as a starting point for understanding NaAlSi. 

\section{Discussion}
The classes of materials that contain relatively high temperature superconductors\cite{other}
continues to expand. Superconductors derived from doped 2D semiconductors pose many of the most
interesting issues in superconductivity today. The cuprates and the Fe-pnictides
(and -chalcogenides) are strongly magnetic, and comprise one end of the spectrum
(though they are themselves quite different). On the other end lie those with
little, perhaps negligible, magnetism: electron-doped ZrNCl and HfNCl, and
electron-doped TiNCl. There are several other, lower-$T_c$ systems, whose
behavior seems different still (hydrated Na$_x$CoO$_2$, Li$_{1-x}$NbO$_2$, and
several transition-metal disulfides and diselenides).

A common feature of most of these systems is that they are 2D and have a small,
but not tiny, concentration of charge carriers, often in the range of 0.05--0.15
carriers per unit cell. These materials also have ionic character. NaAlSi
differs in that it has $sp$ carriers---the others have carriers in $d$ bands---and is
self-doped, a compensated semimetal.
We suggest that a useful view of NaAlSi is that it be regarded as arising from an underlying
ionic semiconductor, but that it has a small {\it negative} gap rather than 
a true gap.  Without the overlap of
the valence and conduction bands, it would be a 2D, ionic, and somewhat
covalent semiconductor like the aforementioned nitridochloride compounds, which
superconduct in the 15--25 K range. 
Comparing the characteristics of these two classes of superconductors
should further the understanding of 2D superconductivity.

\section{Acknowledgments}
This work was supported by DOE grant DE-FG02-04ER46111,
the Strategic Sciences Academic Alliance Program under grant
DE-FG03-03NA00071, and by DOE SciDAC Grant No. DE-FC02-06ER25794.

\end{document}